\documentclass[3p,times]{elsarticle}
\usepackage{ecrc}
\usepackage{epstopdf}
\usepackage{graphicx}
\usepackage{amsmath,amssymb}
\usepackage{subfigure}

\volume{00}
\firstpage{1}
\journalname{Nuclear Physics A}
\runauth{J.~B.~Rose et al.}
\jid{nupha}
\jnltitlelogo{Nuclear Physics A}

\begin{document}

\begin{frontmatter}

\title{Extracting the bulk viscosity of the quark-gluon plasma}

\author[McGill]{Jean-Bernard Rose} 
\author[McGill]{Jean-Francois Paquet} 
\author[McGill]{Gabriel S. Denicol}
\author[McGill,LBNL]{Matthew Luzum}
\author[BNL]{Bjoern Schenke}
\author[McGill]{Sangyong Jeon}
\author[McGill]{Charles Gale}

\address[McGill]{Department of Physics, McGill University, 3600 rue University, Montr\'eal, Qu\'ebec H3A 2T8, Canada}
\address[LBNL]{Lawrence Berkeley National Laboratory, Berkeley, CA 94720, USA}
\address[BNL]{Physics Department, Brookhaven National Lab, Building 510A, Upton, NY, 11973, USA }

\begin{abstract}
We investigate the implications of a nonzero bulk viscosity coefficient on the azimuthal momentum anisotropy of ultracentral
relativistic heavy ion collisions at the Large Hadron Collider. We find that, with IP-Glasma initial conditions,
a finite bulk viscosity coefficient leads to a better description of the flow harmonics in ultracentral collisions. 
We then extract optimal values of bulk and shear viscosity coefficients that provide the best agreement with flow harmonic coefficients data in this centrality class.

\end{abstract}

\end{frontmatter}

\section{Introduction}

\label{intro} Currently, most fluid-dynamical simulations of relativistic
heavy ion collisions take into account only dissipative effects originating
from shear viscosity. However, QCD is a nonconformal field theory and in
principle there is no a priori reason to neglect the effects of bulk viscous
pressure. In this contribution, we explore the phenomenological implications
of a nonzero bulk viscosity coefficient on the azimuthal momentum anisotropy
of ultracentral heavy ion collisions at the Large Hadron Collider (LHC). We
then extract optimal values of bulk and shear viscosity coefficients that
are able to describe the flow harmonic coefficients in this centrality
class. We find that, with IP-Glasma initial conditions, a finite bulk
viscosity coefficient improves the description of flow harmonics in
ultracentral collisions.

\section{Fluid-dynamical model of the collision}

In this work we simulate ultrarelativistic heavy ion collisions in the
0--1\% centrality class at LHC energies. The initial energy density profile
is obtained using the IP-Glasma initial condition model, with a
thermalization time of $\tau _{0}=0.4$ fm \cite{Schenke:2012fw}. The time
evolution of this system is determined using relativistic dissipative fluid
dynamics, with the main equations of motion being the continuity equation
for the energy-momentum tensor, $T^{\mu \nu }$, $\partial _{\mu }T^{\mu \nu
}=0$, where $T^{\mu \nu }=\varepsilon u^{\mu }u^{\nu }-\Delta ^{\mu \nu
}(P_{0}+\Pi )+\pi ^{\mu \nu }$, with $\varepsilon $ being the energy
density, $P_{0}$ the thermodynamic pressure, $\Pi $ the bulk viscous
pressure, and $\pi ^{\mu \nu }$ the shear-stress tensor. We further
introduced the projection operator $\Delta ^{\mu \nu }=g^{\mu \nu }-u^{\mu
}u^{\nu }$ onto the 3-space orthogonal to the fluid velocity. We assume that
the baryon number density and diffusion are zero at all space-time points
and our metric convention is $g^{\mu \nu }=\mathrm{diag}(+1,-1-1-1)$.

The conservation laws and equation of state $P_{0}(\varepsilon )$ are not
enough to determine the time evolution of $T^{\mu \nu }$. In order to do so,
one still needs to provide the time-evolution or constitutive equations
satisfied by $\Pi $ and $\pi ^{\mu \nu }$. For this purpose, we employ
relaxation-time equations derived from kinetic theory \cite%
{DNMR,Denicol:2014vaa}, solved numerically using the \textsc{music}
hydrodynamic simulation \cite{Schenke:2010nt,Marrochio:2013wla}.
Explicitly, we solve%
\begin{eqnarray}
\tau _{\Pi }\dot{\Pi}+\Pi &=&-\zeta \theta -\delta _{\Pi \Pi }\Pi \theta
+\lambda _{\Pi \pi }\pi ^{\mu \nu }\sigma _{\mu \nu }\;,  \label{intro_1} \\
\tau _{\pi }\dot{\pi}^{\left\langle \mu \nu \right\rangle }+\pi ^{\mu \nu }
&=&2\eta \sigma ^{\mu \nu }+2\tau _{\pi }\pi _{\alpha }^{\left\langle \mu
\right. }\omega ^{\left. \nu \right\rangle \alpha }-\delta _{\pi \pi }\pi
^{\mu \nu }\theta +\varphi _{7}\pi _{\alpha }^{\left\langle \mu \right. }\pi
^{\left. \nu \right\rangle \alpha }-\tau _{\pi \pi }\pi _{\alpha
}^{\left\langle \mu \right. }\sigma ^{\left. \nu \right\rangle \alpha
}+\lambda _{\pi \Pi }\Pi \sigma ^{\mu \nu }.  \label{intro_2}
\end{eqnarray}%
The above equations of motion include nonlinear terms that couple bulk
viscous pressure to the shear-stress tensor and vice-versa. Recently, they were
shown to be in good agreement with solutions of the 0+1 Anderson-Witting
equation in the massive limit \cite{Denicol:2014mca}. The transport
coefficients $\tau _{\Pi }$, $\zeta $, $\delta _{\Pi \Pi }$, $\lambda _{\Pi
\pi }$, $\tau _{\pi }$, $\eta $, $\delta _{\pi \pi }$, $\varphi _{7}$, $\tau
_{\pi \pi }$, and $\lambda _{\pi \Pi }$ are complicated functions of
temperature. For strongly coupled fluids, such as the quark-gluon plasma
created at the LHC, their values are not known. As a first step, we fix the parametric
dependence of all transport coefficients on the shear viscosity, $\eta$, and bulk viscosity, $\zeta$, using formulas derived from the Boltzmann
equation near the conformal limit \cite{Denicol:2014vaa}.

The shear viscosity is assumed to be proportional to the entropy density,
i.e., $\eta =as$, where the parameter $a$ is fixed by fitting the data. In
simulations using IP-Glasma initial conditions that include only dissipative
effects from shear viscosity, one usually obtains $a=0.20-0.22$ at LHC
energies \cite{Gale:2012rq,Denicol:2014ywa}. The bulk viscosity coefficient
is parametrized as \cite{Denicol:2014vaa}
\begin{equation}
\zeta =b\times \eta \left( \frac{1}{3}-c_{s}^{2}\right) ^{2},
\label{intro_2}
\end{equation}%
where $c_{s}$ is the velocity of sound and $b$ a parameter that will also be
adjusted to fit the data. Of course, the actual temperature dependence of
each viscosity coefficient is not known for QCD matter. Nevertheless, the
aforementioned parametrizations should be able to capture the relevant
features of dissipation and will be enough for the purposes of this initial
study.

The final state hadrons are produced following the usual Cooper-Frye
procedure \cite{Cooper:1974mv}, in the same way as outlined in Ref.~\cite%
{Denicol:2014ywa}. The only difference is that we have to specify the
nonequilibrium correction to the hadron momentum distribution function due
to bulk viscosity, in addition to the usual correction due to shear
viscosity . Here, we employ the distribution derived from the Boltzmann
equation using the relaxation time approximation, also used by Bozek \textit{%
et al} \cite{Bozek:2009dw},%
\begin{eqnarray}
\delta f_{\Pi }^{i} &=&-f_{0\mathbf{k}}^{i}\left( 1\pm f_{0\mathbf{k}%
}^{i}\right) \left[ \frac{1}{u_{\mu }k_{i}^{\mu }}\frac{m_{i}^{2}}{3}-\left( 
\frac{1}{3}-c_{s}^{2}\right) u_{\mu }k_{i}^{\mu }\right] \frac{\Pi }{C_{\Pi }%
}, \\
C_{\Pi } &=&\frac{1}{3}\sum_{i=1}^{N}m_{i}^{2}g_{i}\int dK_{i}\text{ }f_{0\mathbf{k}}^{i} \left( 1\pm f_{0\mathbf{k}%
}^{i}\right) \left[ \frac{1}{u_{\mu }k_{i}^{\mu }}\frac{%
m_{i}^{2}}{3}-\left( \frac{1}{3}-c_{s}^{2}\right) u_{\mu }k_{i}^{\mu }\right]
.
\end{eqnarray}%
Above, $k_{i}^{\mu }$, $m_{i}$, and $g_{i}$ are the 4-momentum, mass, and
degeneracy factor of the $i$--th hadron species, respectively, $N$ is the
number of hadrons considered in the calculation, and $f_{0\mathbf{k}}^{i}$
is the Fermi-Dirac or Bose-Einstein distribution function.

We note that the IP-Glasma initial conditions used in this work include the
effect of nucleon-nucleon correlations in the initial state. Such
correlations were recently found to be important when simulating central and
ultracentral collisions with the IP-Glasma model \cite{Denicol:2014ywa}.

\section{Results}

We calculate the integrated elliptic flow coefficient $v_{n}\{2\}$ via the
cumulant method. In order to compare with CMS data \cite{CMS:2013bza}, we
use a lower cut off of 0.3 GeV when performing integrals in the transverse
momentum, $p_{T}$. All our calculations were done using the parametrization
of the lattice QCD equation of state by Huovinen and Petreczky \cite%
{Huovinen:2009yb}. The calculations with bulk viscosity were all performed
in chemical equilibrium and with a freezeout temperature of $T_{FO}=140$
MeV. The calculations with only shear viscosity were performed using the
same parameters as in \cite{Denicol:2014ywa}, with a chemical freezeout
temperature of $T_{CF}=150$ MeV and $T_{FO}=103$ MeV. In all figures, the
shaded bands represent the statistical uncertainty.

\begin{figure}[th]
\centering
\subfigure[]{\includegraphics[width=0.35\textwidth]{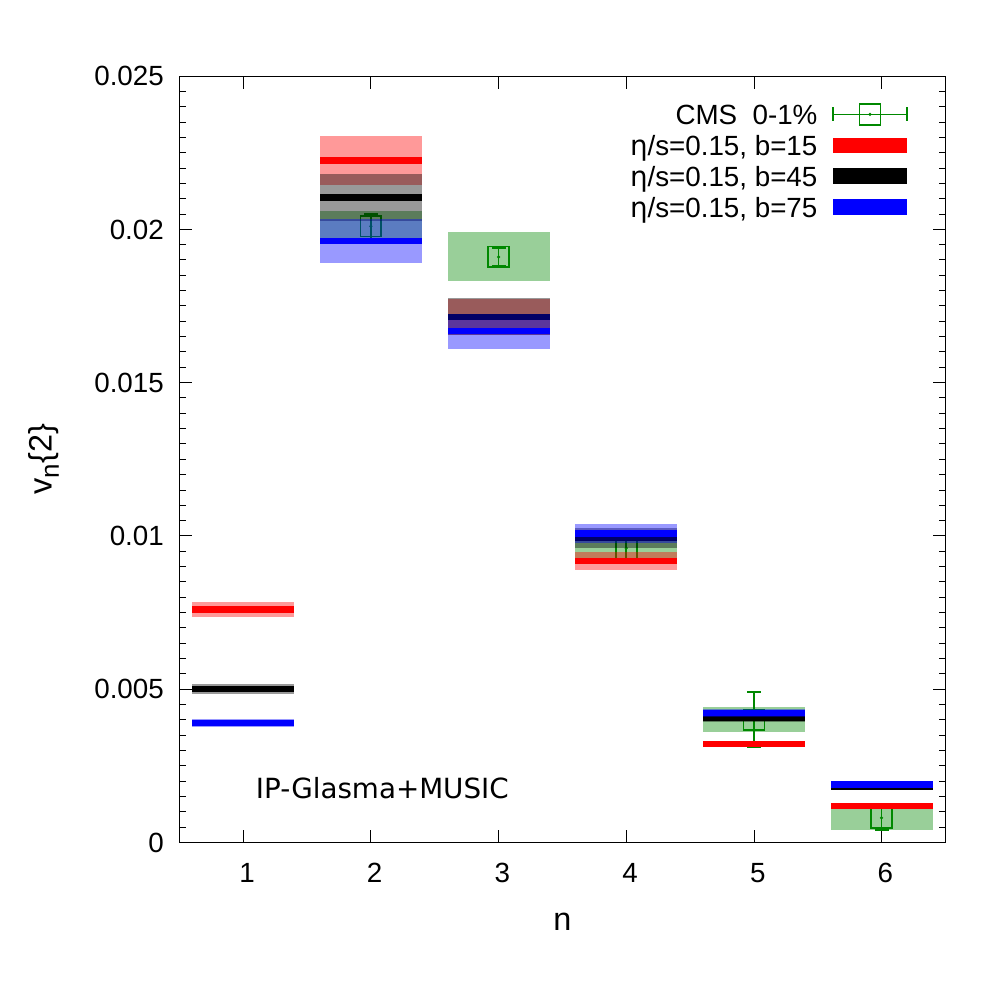}
\label{VnCMS}} \hspace{1.5cm} 
\subfigure[]{\includegraphics[width=0.35\textwidth]{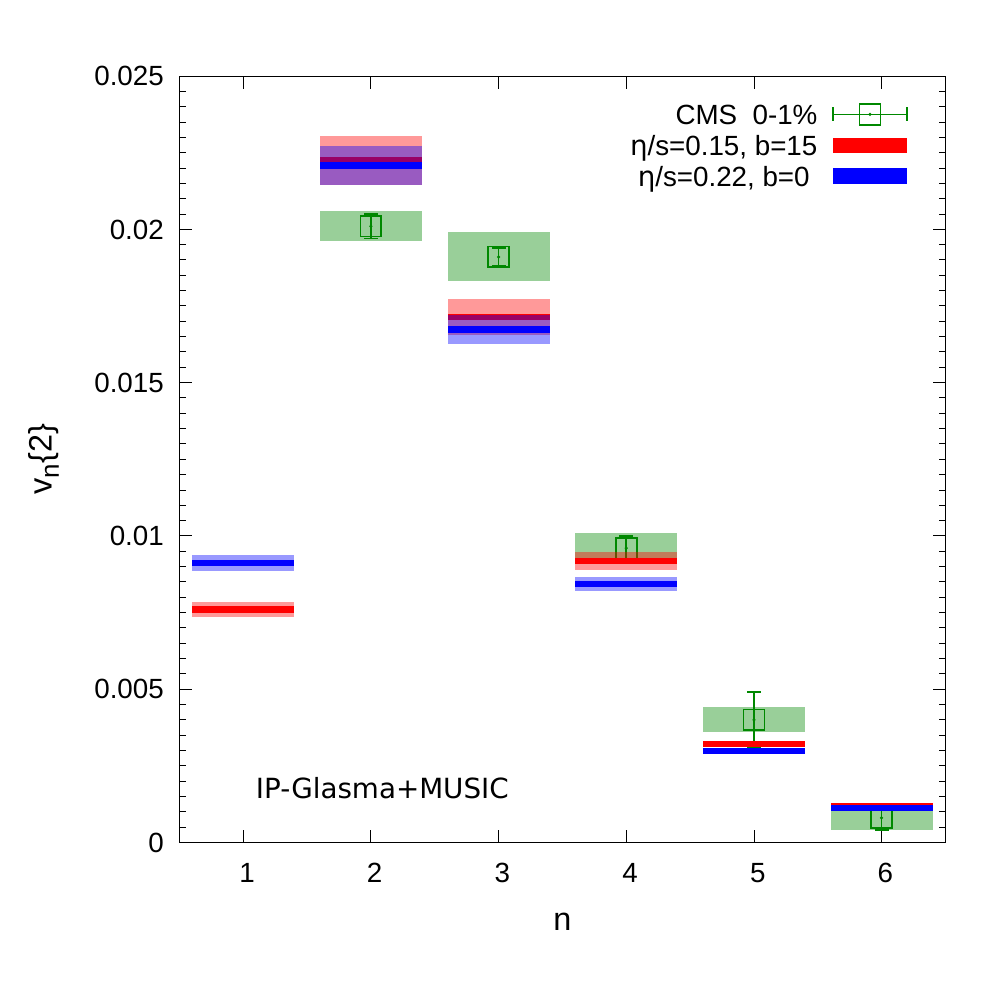}
\label{VnCMS2}}
\caption{ (Color online) Comparison of the flow harmonics, $v_n\{2\} $,
calculated using IP-Glasma initial conditions, with with CMS data for the
0-1\% centrality class and for several values of the parameter b. }
\label{VnCMS}
\end{figure}

In Fig.~\ref{VnCMS} we show the results for a given shear viscosity, $\eta
/s=0.15$, and several values of $b=15$, $45$, and 75. We see that increasing
the value of bulk viscosity improves the agreement with the flow harmonics
measured by CMS. In Fig.~\ref{VnCMS2}, we compare a simulation which
includes only shear viscosity, with $\eta /s=0.22$, to another with $\eta
/s=0.15$ and $b=15$. Both calculations provide an almost equally good
agreement with the flow harmonics measured by CMS, but with rather different
values of shear viscosity. In this case, including bulk viscosity reduced
the value of $\eta /s$ by $32\%$, from 0.22 to 0.15.

\begin{figure}[th]
\centering
\subfigure[]{\includegraphics[width=0.35\textwidth]{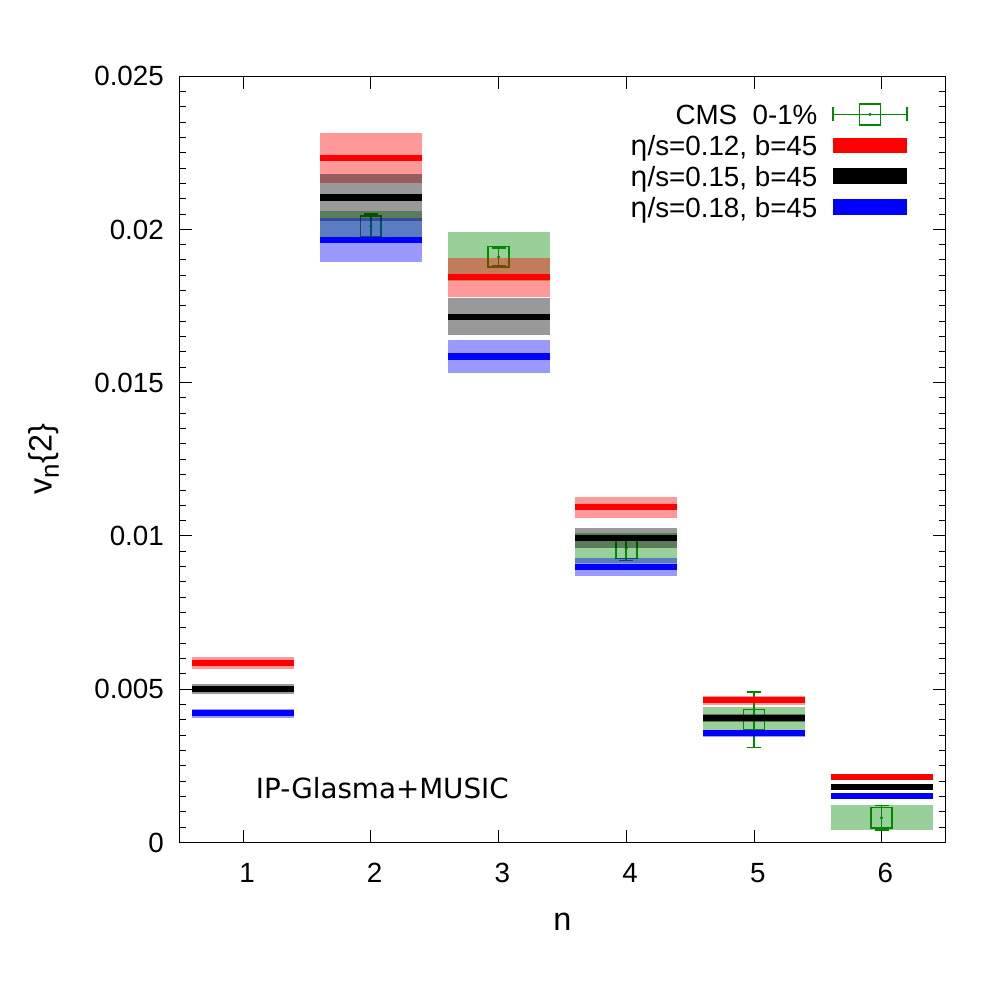}
\label{VnCMS3}} \hspace{1.5cm} 
\subfigure[]{\includegraphics[width=0.35\textwidth]{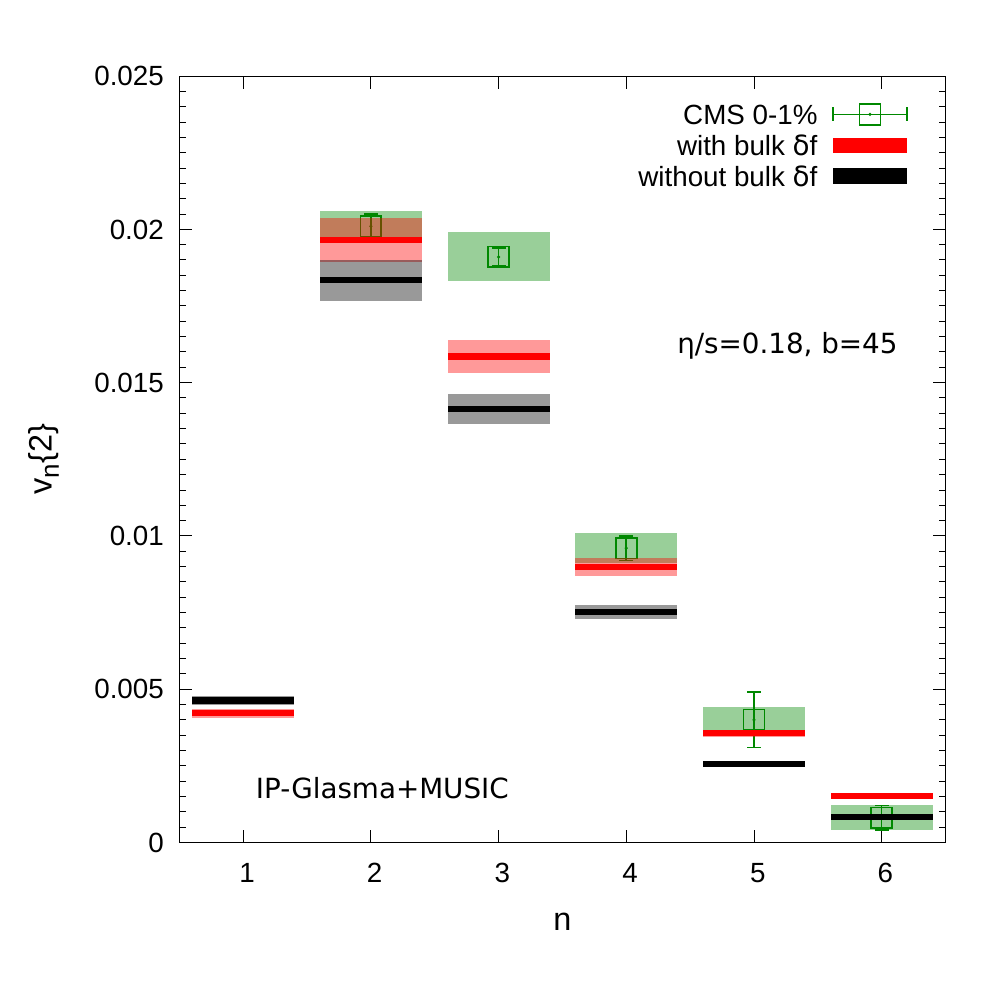}
\label{VnCMS4}}
\caption{ In the left panel we show the flow harmonics, $v_n\{2\} $,
calculated using IP-Glasma initial conditions, for $b=45$ and several values
of $\protect\eta/s$ (left panel). In the right panel we show the effect of $%
\protect\delta^{i}f_{\Pi }$ on $v_n\{2\} $. }
\end{figure}

In Fig.~\ref{VnCMS3} we show what happens when the parameter $b$ is fixed as 
$b=45$ and the shear viscosity is varied, taking the values $\eta /s=0.12$,
0.15, and 0.18. Note that the parametrization of bulk viscosity used in this
work, Eq.~(\ref{intro_2}), is linear in the shear viscosity, so keeping $b$
constant does not imply that the bulk viscosity is kept fixed. In this figure we see
that the value of shear viscosity that provides the best overall agreement
with the CMS data is $\eta /s=0.15$, which was the value already employed in
the simulations discussed above.

In Fig.~\ref{VnCMS4} the results for $\eta /s=0.18$ and $b=$45 are shown
calculated with and without the bulk component of $\delta f$, $\delta
^{i}f_{\Pi }$. In contrast to the shear correction, the bulk correction to
the single particle distribution function used in this work lead to an
increase in the flow harmonics. Such an increase was systematically stronger
for higher harmonics: $v_{2}\{2\}$ is increased by 4\%, $v_{3}\{2\}$ by 8\%, 
$v_{4}\{2\}$ by 20\%, $v_{5}\{2\}$ by 57\%, and $v_{6}\{2\}$ by 117\%.
Therefore, in order to properly describe higher harmonics, it is important
to take into account the $\delta ^{i}f_{\Pi }$ contribution. However, one
should note that different momentum dependences for this quantity can
provide rather different results.

\begin{figure}[th]
\centering
\includegraphics[width=0.45\textwidth]{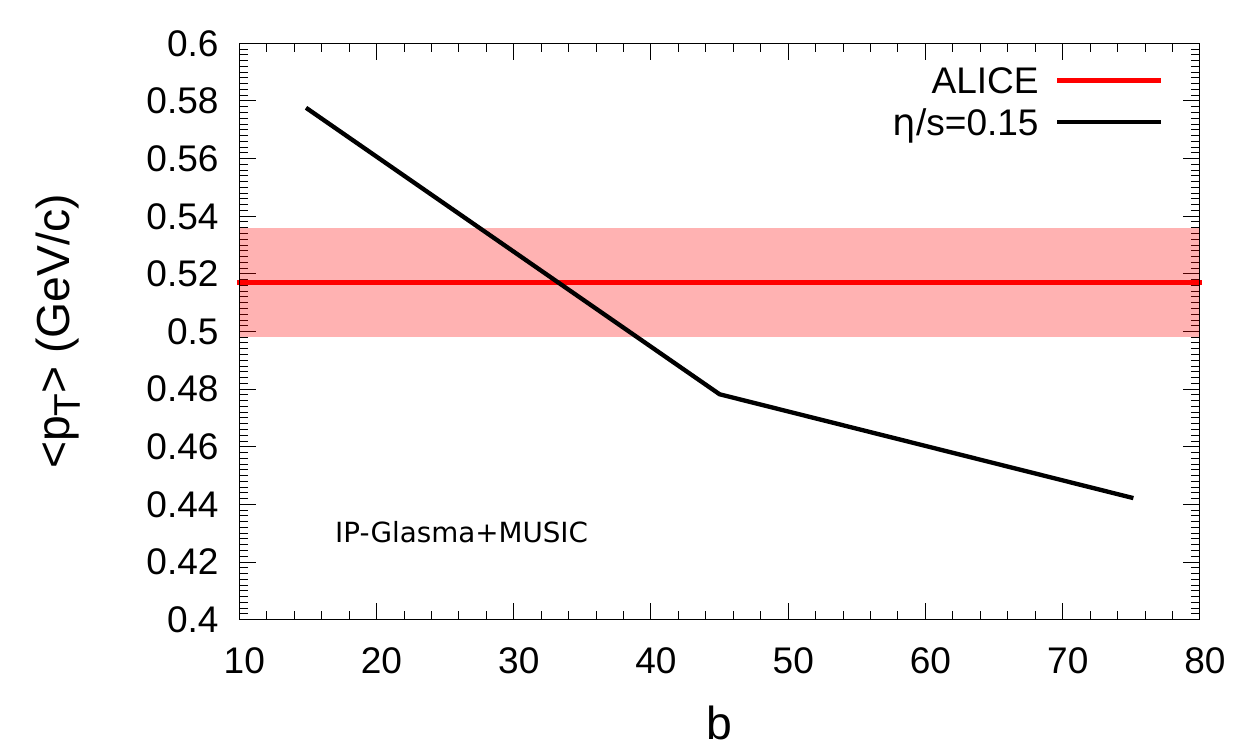}
\caption{ (Color online) The mean transverse momentum of pions, $\langle
p_{T}^{\protect\pi } \rangle$, for several values of $b$ compared to the
ALICE data \protect\cite{Abelev:2013vea}.}
\label{VnCMS5}
\end{figure}

Finally, we show the effect of bulk viscosity on the mean transverse
momentum of pions, $\langle p_{T}^{\pi } \rangle$, in Fig.~\ref{VnCMS5}. Our
calculations are compared to ALICE data \cite{Abelev:2013vea} in the 0-5\% centrality class and, for this
purpose, we used a lower cut-off of 0.12 GeV when performing integrals in
transverse momentum. The effect of bulk viscosity on  $\langle p_{T}^{\pi } \rangle$ is not
small and can be used to further restrict the size of the bulk viscosity
coefficient. Here, we see that the best description of $\langle p_{T}^{\pi }
\rangle$ happens when $b=$20--40, for the case where $\eta /s=0.15$. Note
that the $\langle p_{T}^{\pi } \rangle$ has a strong dependence on $\delta
^{i}f_{\Pi }$, so the optimal values we obtained can be modified if another $%
\delta ^{i}f_{\Pi }$ is used.

\section{Conclusions}

In this contribution we performed a preliminary study on the effects of bulk
viscous pressure in ultracentral relativistic heavy ion collisions at the
LHC. We found that bulk viscous pressure can have non-negligible effect on
the flow harmonic coefficients in this centrality class. Furthermore, we
showed that, at least for IP-Glasma initial conditions, the inclusion of
bulk viscosity leads to a better description of the data. Although describing the
anisotropy distribution of ultracentral events remains a challenge, we
expect bulk viscosity to be a key ingredient in understanding such observables.

%\section{Acknowledgments}

\textbf{Acknowledgments:} The authors thank H.~Niemi and J.~Jia for fruitful discussions. This work
was supported in part by the Natural Sciences and Engineering Research
Council of Canada, and by the U. S. DOE Contract No. DE-AC02-98CH10886.
G.S.~Denicol acknowledges support through a Banting Fellowship of the
Natural Sciences and Engineering Research Council of Canada, and C.~G.~
acknowledges support from the Hessian Initiative for Excellence (LOEWE)
through the Helmholtz International Center for FAIR (HIC for FAIR).

{}

\end{document}